\documentclass[journal,12pt,draftclsnofoot,onecolumn]{IEEEtran}

\usepackage[cmex10]{amsmath}
\usepackage{amssymb}

\usepackage{cases}
\usepackage{multirow}
\usepackage{empheq}

\usepackage{cite}
\usepackage{verbatim}

\usepackage{algorithm}
\usepackage{algpseudocode}

\usepackage{siunitx}

\ifCLASSINFOpdf
\usepackage[pdftex]{graphicx}
\else
\usepackage[dvips]{graphicx}
\fi
\usepackage[caption=false,font=footnotesize]{subfig}

\usepackage{color}
\usepackage{multirow}

\newtheorem{theorem}{Theorem}
\newtheorem{lemma}[theorem]{Lemma}

\newtheorem{definition}[theorem]{Definition}

\newtheorem{proposition}[theorem]{Proposition}
\newtheorem{remark}[theorem]{Remark}

\newcommand{\vect}[1]{\mathbf{#1}}

\begin{document}
	
\bstctlcite{IEEEexample:BSTcontrol}
	
	\sloppy
	
	\title{Optimizing Write Fidelity of MRAMs via \\ Iterative Water-filling Algorithm}
	
	
	\author{\IEEEauthorblockN{Yongjune Kim,
			Yoocharn Jeon, Hyeokjin Choi, 	
			Cyril Guyot, and 
			Yuval Cassuto}
	\thanks{
		This work was presented in part at the IEEE International Symposium on Information Theory, Los Angeles, CA, USA, June 2020~\cite{Kim2020optimizing}.
	}
	\thanks{	
		Y. Kim and H. Choi are with the Department of Information and Communication Engineering, Daegu Gyeongbuk Institute of Science and Technology (DGIST), Daegu 42988, South Korea (e-mail: yjk@dgist.ac.kr). Y. Jeon and C. Guyot are with Western Digital Research, Milpitas, CA 95035 USA. Y. Cassuto is with the Viterbi Department of Electrical and Computer Engineering, Technion -- Israel Institute of Technology, Haifa, Israel.}
	}

	
	\maketitle
	
	\begin{abstract}
		Magnetic random-access memory (MRAM) is a promising memory technology due to its high density, non-volatility, and high endurance. However, achieving high memory fidelity incurs significant write-energy costs, which should be reduced for large-scale deployment of MRAMs. In this paper, we formulate a \emph{biconvex} optimization problem to optimize write fidelity given energy and latency constraints. The basic idea is to allocate non-uniform write pulses depending on the importance of each bit position. The fidelity measure we consider is mean squared error (MSE), for which we optimize write pulses via \emph{alternating convex search (ACS)}. By using Karush-Kuhn-Tucker (KKT) conditions, we derive analytic solutions and propose an \emph{iterative water-filling-type} algorithm by leveraging the analytic solutions. Hence, the proposed iterative water-filling algorithm is computationally more efficient than the original ACS while their solutions are identical. Although the original ACS and the proposed iterative water-filling algorithm do not guarantee global optimality, the MSEs obtained by the proposed algorithm are comparable to the MSEs by complicated global nonlinear programming solvers. Furthermore, we prove that the proposed algorithm can reduce the MSE exponentially with the number of bits per word. For an 8-bit accessed word, the proposed algorithm reduces the MSE by a factor of 21. We also evaluate the proposed algorithm for MNIST dataset classification supposing that the model parameters of deep neural networks are stored in MRAMs. The numerical results show that the optimized write pulses can achieve \SI{40}{\%} write energy reduction for a given classification accuracy. 
	\end{abstract}
	
	\section{Introduction}
	
	Magnetic random access memory (MRAM) is a nonvolatile memory technology that has a potential to combine the speed of static RAM (SRAM) and the density of dynamic RAM (DRAM). MRAM technology is attractive since it provides high endurance and complementary metal-oxide-semiconductor (CMOS) compatibility~\cite{Zhu2008magnetoresistive,Kim2015spin}. 
	
	In spite of its attractive features, one of the main challenges is the high energy consumption to write information \emph{reliably} in the memory element~\cite{Zhu2008magnetoresistive,Kim2015spin,Kim2012write}. In an MRAM device, a memory state ``1'' or ``0'' is determined by the magnetic moment orientation of the memory element~\cite{Zhu2008magnetoresistive}. Switching the magnetic moment orientation requires high write-current, which introduces write errors when the energy budget is limited~\cite{Kim2015spin}. In addition, high current injection through the tunneling barriers incurs a severe stress and leads to breakdown, which degrades the endurance of MRAM cells~\cite{Kim2012write,Khvalkovskiy2013basic}. Hence, one of the key directions of MRAM research has been toward providing reliable switching with limited energy cost. At the device level, new materials~\cite{Ikeda2010perpendicular,Meng2006spin} or new switching mechanisms~\cite{Nozaki2010voltage,Wang2012electric} have been explored. Several architectural techniques~\cite{Kim2012write,Zhou2009energy,Ranjan2015approximate} to reduce write energy have been investigated. 
	
	However, prior efforts have not considered the differential importance of each bit position in error tolerant applications such as signal processing and machine learning (ML) tasks. In these applications, the impact of bit errors depends on the bit position, i.e., most significant bits (MSBs) are more important than least significant bits (LSBs)~\cite{Mittal2016survey,Alioto2017energy}. This differential importance has been leveraged to effectively optimize energy in major memory technologies such as SRAMs~\cite{Frustaci2016approximate,Yang2011unequal,Kim2018generalized} and DRAMs~\cite{Cho2014edram,Kim2019optimal}. These works attempt to minimize the mean squared error (MSE) since the MSE is a more meaningful fidelity metric than the write-failure probability (or bit error rate) in error tolerant applications. 
	
	In this paper, we provide a \emph{principled} approach to improving MRAM's write fidelity. We formulate a \emph{biconvex optimization} problem to minimize the MSE for given write energy and latency constraints. Since the objective function (i.e., MSE) and the constraints (write energy and latency) depend on the write-current and the write-duration, we attempt to optimize both parameters by solving the biconvex problem.     
	
	A biconvex problem is an optimization problem where the objective function and the constraint set are biconvex~\cite{Gorski2007biconvex}, which, roughly speaking, means convexity in one subset of the variables when the variables in the complement set are fixed, and vice versa. A common algorithm for solving biconvex problems is \emph{alternating convex search (ACS)} (or \emph{alternating convex optimization (ACO)}), which updates each variable subset by fixing another and solving the corresponding convex problem in an iterative manner~\cite{Wendell1976minimization}. ACS or its variants are widely used to solve biconvex optimization problems in location theory~\cite{Cooper1963location}, image processing~\cite{Besl1992method}, and machine learning (e.g., non-negative matrix factorization~\cite{Lee2001algorithms}). 
	
	We adopt ACS approach to optimize the write-current and the write-duration. By taking into account unique properties of MRAM's write operations,  we derive analytic solutions of the formulated biconvex problem. Since each iteration of the ACS algorithm corresponds to solving a convex problem, we can derive the analytic solutions based on the Karush-Kuhn-Tucker (KKT) conditions. Then, by leveraging these analytic solutions for each iteration, we propose an \emph{iterative water-filling algorithm} that is more efficient than the original ACS.         
	
	In general, ACS-type algorithms cannot guarantee the global optimal solution since biconvex problems may have a large number of local minima~\cite{Gorski2007biconvex}. We prove that the proposed iterative water-filling algorithm converges and reduces the MSE exponentially. Moreover, numerical results show that the proposed algorithm achieves results comparable to complicated nonlinear programming solvers such as the mesh adaptive direct search (MADS) algorithm~\cite{Audet2006mesh,LeDigabel2011algorithm}. Also, we evaluate MNIST dataset classification by supposing that the model parameters of deep neural networks are stored in MRAM cells. Numerical results show that the optimized write-energy allocation by the proposed algorithm can achieve \SI{40}{\%} write energy reduction to achieve the same classification accuracy. 
	
	Originally, iterative water-filling algorithms were proposed for multi-user communications settings, e.g., interference and multiple-access channels. For a Gaussian vector multiple-access channel, the authors of~\cite{Yu2004iterative} proposed an iterative water-filling-type algorithm to maximize the sum capacity. It decomposes the multiuser problem into a sequence of single-user problems. The MRAM iterative water-filling algorithm proposed here decomposes the write optimization problem into a sequence of write-current optimization problem and write-duration optimization problem. Unlike the algorithm in \cite{Yu2004iterative} where the single-user problems are identical (only with different channel and noise parameters), our algorithm solves two different problems of write-current and write-duration assignments. Also, we note that our optimization problem is non-convex whereas maximizing the sum capacity of Gaussian vector multiple-access channel is a convex problem.  
	
	Prior optimization studies on voltage swings of SRAMs~\cite{Kim2018generalized} and refresh operations of DRAMs~\cite{Kim2019optimal} are similar in spirit, viz. minimizing the MSE for given resource constraints. However, the MRAM write optimization of this work is \emph{non-convex} whereas the formulated problems in \cite{Kim2018generalized,Kim2019optimal} are convex. Also, we propose the iterative water-filling algorithm and analyze convergence and improvement of the optimized MSE. To the best of our knowledge, our work is the first principled approach to optimization for MRAMs' write operations via a communication theoretic approach. 
	
	The rest of this paper is organized as follows. Section~\ref{sec:mram} explains the basics of MRAM and the challenges of high write energy consumption. Section~\ref{sec:model} introduces the optimization metrics for MRAM write operations. Section~\ref{sec:optimization} formulates the biconvex problem to optimize the write-current and write-pulse assignments. Section~\ref{sec:analysis} proposes the iterative water-filling algorithm by leveraging analysis on the optimized solutions. Section~\ref{sec:convergence} shows convergence and MSE reduction of the proposed algorithm. Section~\ref{sec:numerical} provides numerical results and Section~\ref{sec:conclusion} concludes. 
	
	\section{Basic Principles of MRAMs}\label{sec:mram}
	
	\subsection{Basics of MRAMs}

\begin{figure}[t]
		\centering
		\includegraphics[width=0.35\textwidth]{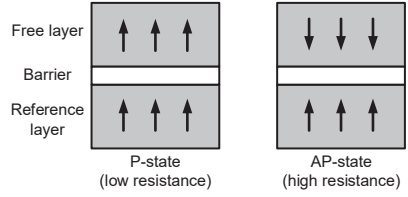}
			\vspace{-4mm}
		\caption{P state and AP state of MTJ MRAM devices.}
		\label{fig:state}
		\vspace{-4mm}
	\end{figure}

	MRAM cells store information by controlling bistable magnetization of ferromagnetic material and retrieve information by sensing resistance of magnetic tunnel junctions (MTJs). An MTJ device consists of two ferromagnetic layers of reference layer (RL) and free layer (FL), separated by a  thin tunneling barrier. RL has a very stable magnetization and it maintains the magnetization throughout all operations, while FL can be switched between two stable magnetization states by a moderate stimulus. The resistance of an MTJ depends on the relative orientation of the FL magnetization with respect to that of the RL as shown in Fig.~\ref{fig:state}. If the magnetizations of FL and RL are in the same direction (parallel- or P-state), then the corresponding resistance is low. The opposite direction (antiparallel- or AP-state) results in high resistance. The difference in tunneling currents between a P-state (low resistance) and a AP-state (high resistance) is utilized to encode binary data~\cite{Zhu2008magnetoresistive,Kim2015spin}.
	
	Writing information into an MTJ is performed by driving a sufficient current through it. Depending on the current's direction, one can flip the magnetization of the FL into P- or AP-state. If a current flows from FL to RL (electrons from RL to FL), electrons are spin-polarized along the magnetization of RL while passing through the layer. The electrons transmitted from the RL interact and exchange the magnetic moments with ones in the FL. If the MTJ is in the AP-state and the current is sufficiently high, then the magnetization orientation is flipped to P-state. When the current is reversed, incoming electrons are polarized along the magnetization of FL. Since the RL's magnetization is parallel to the FL, the majority of the electrons tunnel the barrier while the minority that have antiparallel magnetizations are reflected. Because of this selective tunneling, the antiparallel spins are accumulated in the FL. If the enriched antiparallel spin dominates the FL, it flips the magnetization of the FL into the AP-state.

	The magnetization switching between P state and AP state is not deterministic. The write (switching) failure probability depends on the current  and duration of the write pulse as follows~\cite[Eq. (26)]{Khvalkovskiy2013basic}: 
	\begin{align}\label{eq:wfp}
	p_{\text{WF}}(i,t) = 1 - \exp\left(-\frac{\Delta \pi^2 (i-1)}{4\left\{i \exp(2(i-1)t) - 1 \right\}}\right) 
	\end{align}
	where $\Delta$ denotes the thermal stability factor. The normalized current $i$ and the normalized duration $t$ are given by
	\begin{align}\label{eq:fabrication}
		i = \frac{I}{I_{c}}, \quad t = \frac{T}{T_c} 
	\end{align}
	where $I$ denotes the actual write-current and $I_{c}$ is the critical current. Also, $T$ denotes the actual write-duration and $T_c$ is the characteristic relaxation time. Note that $\Delta$, $I_{c}$, and $T_c$ are fabrication parameters~\cite{Khvalkovskiy2013basic,Butler2012switching}.
	
	To ensure a low failure probability, we should control the current or the duration judiciously. A longer write-duration may lower the write-failure probability at the expense of longer write latency and higher energy consumption. Instead of increasing the write-duration, we can adopt higher write-current. However, it increases the write energy and the risk of dielectric breakdown of the MTJ.
	
	\subsection{Subarray Architecture}
		
	\begin{figure}[t]
		\centering
		\includegraphics[width=0.46\textwidth]{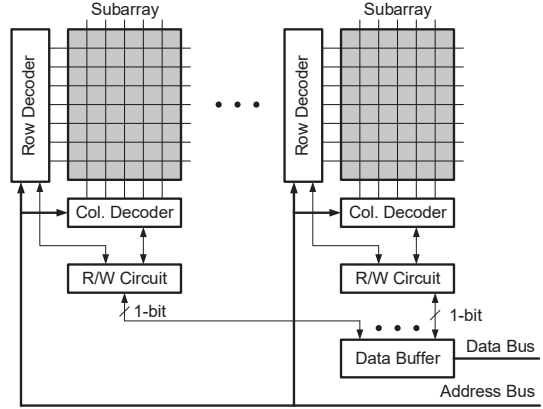}
		\vspace{-4mm}
		\caption{MRAM subarray architecture where each subarray consists of $n_{\text{row}}$ rows and $n_{\text{col}}$ columns.}
		\label{fig:arch}
		\vspace{-4mm}
	\end{figure}

	The MRAM cells are arranged in arrays and each of the cells is selectively connected to the read/write circuits to access the data. The metal-oxide-semiconductor field-effect transistors (MOSFETs) are commonly used for the selectors in DRAMs where the required current for memory operations is low enough; a MOSFET with minimum feature sizes can drive the required current. However, the required MRAM write-current is more than an order of magnitude higher than that of DRAMs, which requires MOSFETs with large channel width to drive high write-current. They are not suitable for high-density memories because of large area on a silicon substrate. 
	
	In order to handle this problem, each MRAM cell consists of an MTJ and a threshold switching selector~\cite{Kim2015spin,Yang2017threshold}. These MRAM cells are populated in a crossbar array. To access an MRAM cell, a voltage higher than the threshold voltage of the selector is applied, which turns on the corresponding selector between the selected row-line and column-line, while all the unselected row-lines and column-lines are biased to a midpoint voltage, which keeps all the unselected cells in the array under biases below their threshold voltages. In this manner, the number of needed MOSFETs driving high currents can be reduced from $n_{\text{row}} \times n_{\text{col}}$ to $n_{\text{row}} + n_{\text{col}}$ for a subarray, which is much better suited for high density memories. 


	Because of the limited current drivability of the row line and the column line drivers, \emph{only one cell can be accessed at a time in each subarray} unlike DRAMs where a whole page (row-line) can be read/written together (see Fig.~\ref{fig:arch}). Multiple subarrays are operated in parallel to match the required data bandwidth. This MRAM architecture provides an opportunity to write each bit by using different write-currents and write-durations.	
		
	\section{Metrics for MRAM Write Operations}\label{sec:model}

	The write-failure probability expression $p_{\text{WF}}(i,t)$ of \eqref{eq:wfp} is too complicated to formulate an optimization problem. Therefore, we define a proxy for $p_{\text{WF}}(i,t)$. 
	
	\begin{definition}\label{def:proxy}The proxy $\widetilde{p}_{\text{WF}}(i,t)$ is given by
		\begin{align} 
			\widetilde{p}_{\text{WF}}(i,t) = c \exp\left(-2(i-1) t\right). \label{eq:proxy}
		\end{align}
	where $c = \frac{\pi^2}{4} \cdot \Delta$.
	\end{definition}
		
	\begin{proposition}\label{lem:proxy}$p_{\text{WF}}(i,t)$ converges to $\widetilde{p}_{\text{WF}}(i,t)$ as $i$ and $t$ increase. 
	\end{proposition}
	\begin{IEEEproof}
		The proof is given in Appendix~\ref{pf:proxy}.
	\end{IEEEproof}
	Fig.~\ref{fig:p_approx} shows that the proxy $\widetilde{p}_{\text{WF}}(i,t)$ is very close to $p_{\text{WF}}(i,t)$, especially for lower $p_{\text{WF}}(i,t)$. 

	\begin{remark} \label{rem:fabrication}
		The thermal stability factor $\Delta$ (i.e., fabrication parameter) affects only the coefficient $c = \frac{\pi^2}{4}\Delta$. The optimization parameters $(i, t)$ determine the exponent. It is favorable since the thermal stability $\Delta$ factor does not affect the optimization of $(i,t)$.
	\end{remark}

	\begin{remark} \label{rem:ber} If random data is written, then the probability of bit error $p(i,t)$ becomes
		\begin{align}\label{eq:ber}
			p(i,t)  = \frac{1}{2} \cdot p_{\text{WF}}(i,t) \simeq \frac{1}{2} \cdot \widetilde{p}_{\text{WF}}(i,t). 
		\end{align}
	Suppose that a binary value $x$ should be updated to a binary value $x'$ for a given memory cell. If $x = x'$, then the write failure does not incur a bit error. Since $x$ and $x'$ are random data, $\Pr(x = x') = \frac{1}{2}$ leads to \eqref{eq:ber}. 
	\end{remark}

	\begin{figure}[t]
		\centering
		\includegraphics[width=0.45\textwidth]{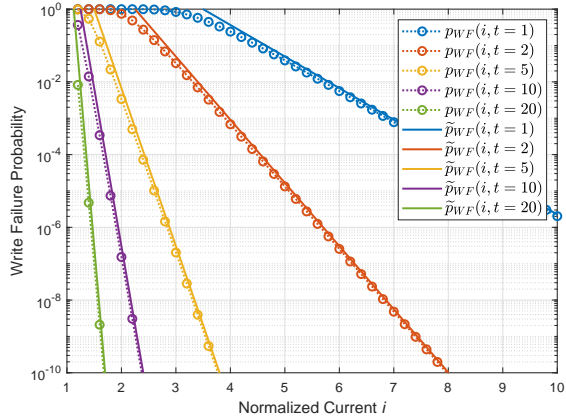}
		\caption{Comparison of the write-failure probability $p_{\text{WF}}(i,t)$ and its proxy $\widetilde{p}_{\text{WF}}(i,t)$ ($\Delta = 60$ as in \cite[Fig. 13]{Khvalkovskiy2013basic}).}
		\label{fig:p_approx}
		\vspace{-4mm}
	\end{figure}

	The normalized energy for writing a single-bit is given by
	\begin{equation} \label{eq:energy_bit}
	\mathsf{E}(i,t) = i^2 t. 
	\end{equation} 
	
	As shown in \eqref{eq:proxy} and \eqref{eq:energy_bit}, the write-current $i$ and the write-duration $t$ are key factors to control the trade-off between write-failure probability and the write energy. If we allocate different write-currents and durations depending on the importance of each bit position, then the corresponding current and duration assignments are given by
	\begin{align}
	\vect{i} = (i_0, \ldots, i_{B-1}),\quad 	\vect{t} &= (t_0, \ldots, t_{B-1})
	\end{align}
	where $i_0$ and $t_0$ define the write pulse for the least significant bit (LSB) and $i_{B-1}$ and $t_{B-1}$ are the write pulse parameters for the most significant bit (MSB). 		


	We define metrics for energy, latency, and fidelity for writing a $B$-bit word. 
	
	\begin{definition}[Energy] \label{def:energy}
		The energy of writing a $B$-bit word is given by
		\begin{equation}
		\mathsf{E}(\vect{i},\vect{t}) = \sum_{b=0}^{B-1}{i_b^2 t_b}. 
		\end{equation}
	\end{definition}

	\begin{definition}[Latency] \label{def:latency}
		The latency of writing a $B$-bit word depends on the maximum write-duration among $\vect{t}=(t_0, \ldots, t_{B-1})$, i.e., 
		\begin{equation}
		\mathsf{L}(\vect{t}) = \max \{t_0, \ldots, t_{B-1}\}. 
		\end{equation}		
	\end{definition}

	Note that $\mathsf{E}(\vect{i},\vect{t})$ and $\mathsf{L}(\vect{t})$ are resource metrics. As a fidelity metric, we consider mean squared error (MSE). 
	
	\begin{definition}[MSE] \label{def:mse}
		The MSE of $B$-bit words is given by
		\begin{align}
		\mathsf{MSE}(\vect{i},\vect{t}) 
		&= \sum_{b=0}^{B-1}{4^b p(i_b, t_b)} \label{eq:mse0} \\
		&\simeq c' \cdot \sum_{b=0}^{B-1}{4^b \exp\left(-2(i_b-1) t_b\right)} \label{eq:mse}
		\end{align} 
		where $c' =\frac{c}{2}$ and the weight $4^b$ represents the differential importance of each bit position. The derivation of \eqref{eq:mse0} can be found in~\cite{Yang2011unequal,Kim2018generalized},  and \eqref{eq:mse} follows from \eqref{eq:ber}. 
	\end{definition}

	\begin{table}[!t]
		\renewcommand{\arraystretch}{1.5}
		\caption{Resource and Fidelity Metrics for Write Operation}
		\vspace{-2mm}
		\label{tab:comparison}
		\centering
		\begin{tabular}{|c|c|c|}	\hline
			 & Metrics & Remarks \\ \hline \hline
			Energy      &  $\mathsf{E}(\vect{i},\vect{t}) = \sum_{b=0}^{B-1}{i_b^2 t_b}$ & Definition~\ref{def:energy} \\ \hline
			Latency     & $\mathsf{L}(\vect{t}) = \max \{t_0, \ldots, t_{B-1}\}$  & Definition~\ref{def:latency} \\ \hline
			Fidelity    & $\mathsf{MSE}(\vect{i},\vect{t}) = \sum_{b=0}^{B-1}{4^b p(i_b, t_b)} $ & Definition~\ref{def:mse} \\ \hline			
		\end{tabular}
	\end{table} 
	
	Table~\ref{tab:comparison} summarizes the defined metrics for writing a $B$-bit word. We note that the fabrication parameters $\Delta$, $I_c$, and $T_c$ do not affect our optimization. Hence, we can reuse the optimized solutions for differently fabricated MRAM devices. 
	
	\section{Optimizing Parameters of Write Operations}\label{sec:optimization}
	
	In this section, we investigate optimization of write-operation parameters. First, the optimized current and duration for single-bit write operations will be investigated and then we formulate a biconvex optimization problem for multi-bit word write operations. 
	
	\subsection{Optimized Parameters for Single-Bit Write Operation}
	
	First, we note that the normalized current should be greater than 1 for a successful write in \eqref{eq:proxy}. It shows that the write-current should be greater than the critical current (i.e., $I > I_c$) so as to switch the direction of magnetization~\cite{Khvalkovskiy2013basic,Butler2012switching}. Then, we can formulate the following optimization problem for single-bit (also multi-bit uniform) write: 
	\begin{equation}
	\begin{aligned} \label{eq:min_wfp}
	& \underset{i, t}{\text{minimize}}
	& & c \exp\left(-2(i-1) t \right)  \\
	&{\text{subject~to}} & & i^2 t \le \mathcal{E}, \quad i \ge 1 + \epsilon, \quad t \ge 0,
	\end{aligned}
	\end{equation} 
	where $\mathcal{E}$ is a constant corresponding to the given write energy budget. We introduce $\epsilon > 0$ to guarantee $i > 1$. 
	This optimization problem is equivalent to  
	\begin{equation}
	\begin{aligned} \label{eq:min_wfp_qe}
	& \underset{i, t}{\text{maximize}}
	& & (i-1) t   \\
	&{\text{subject~to}} & & i^2 t \le \mathcal{E}, \quad i \ge 1 + \epsilon, \quad t \ge 0.
	\end{aligned}
	\end{equation}
	Note that the objective function $(i-1)t$ is not concave. However, we can readily obtain the optimal $i^*$ and $t^*$ as follows.    
	
	\begin{lemma} \label{thm:min_wfp}
		The optimized current and duration for single-bit write are $i^* = 2$ and $t^* = \frac{\mathcal{E}}{4}$, respectively. The proxy (in Definition~\ref{def:proxy}) is given by
		\begin{equation}\label{eq:p_opt}
			\widetilde{p}_{\text{WF}}(i^*, t^*) = c \exp\left(- \frac{\mathcal{E}}{2}\right). 
		\end{equation} 
	\end{lemma}
	\begin{IEEEproof}
		The proof is given in Appendix~\ref{pf:min_wfp}.
	\end{IEEEproof}
	Note that the write-failure probability is an exponentially decaying function of $\mathcal{E}$.  
		
	\subsection{Optimized Parameters for Multi-bit Word Write Operation}
			
	We formulate an optimization problem to determine the write-pulse parameters to write a $B$-bit word. For a given write-energy constraint, we seek to minimize MSE as follows.  
	

	\begin{equation}
	\begin{aligned} \label{eq:min_mse}
	& \underset{\vect{i}, \vect{t}}{\text{minimize}}
	& & \sum_{b=0}^{B-1}{4^b \exp(-2(i_b - 1)t_b)}  \\
	&{\text{subject~to}} & &  \sum_{b=0}^{B-1}{i_b^2 t_b} \le \mathcal{E} \\
	& & & i_b \ge 1 + \epsilon, \quad b=0,\ldots,B-1 \\
	& & & t_b \ge 0, \quad b=0,\ldots,B-1 \\
	\end{aligned}
	\end{equation}
	where the constraints on $i_b$ and $t_b$ are imposed as in \eqref{eq:min_wfp}. We may include an additional latency constraint $\mathsf{L}(\vect{t}) \le \delta$ to guarantee a required write speed performance. 
	
	Although the optimization problem \eqref{eq:min_mse} is not convex, we show that it is a \emph{biconvex} problem.
  
	\begin{definition}[Biconvex Set~\cite{Gorski2007biconvex}] Let $S \subseteq X \times Y$ where $X \subseteq \mathbb{R}^n $ and $Y \subseteq \mathbb{R}^m$ denote two non-empty and convex sets. The set $S$ is defined as a \emph{biconvex set} on $X \times Y$,  if for every fixed $\vect{x} \in X$, $S_\vect{x} \triangleq \left\{\vect{y} \in Y \mid (\vect{x},\vect{y}) \in S \right\}$ is a convex set in $Y$ and for every fixed $\vect{y} \in Y$, $S_\vect{y} \triangleq \left\{\vect{x} \in X \mid (\vect{x},\vect{y}) \in S \right\}$ is a convex set in $X$.
	\end{definition}
	
	\begin{definition}[Biconvex Function~\cite{Gorski2007biconvex}] A function $f:S \rightarrow \mathbb{R}$ is defined as a \emph{biconvex function} on $S$, if for every fixed $\vect{x} \in X$, $f_\vect{x}(\cdot) \triangleq f(\vect{x},\cdot):S_\vect{x} \rightarrow \mathbb{R}$ is a convex function on $S_\vect{x}$, and for every fixed $\vect{y} \in Y$, $f_\vect{y}(\cdot) \triangleq f(\cdot, \vect{y}):S_\vect{y} \rightarrow \mathbb{R}$ is a convex function on $S_\vect{y}$. 
	\end{definition}
		
	\begin{definition}[Biconvex Problem~\cite{Gorski2007biconvex}] An optimization problem of the following form:
		\begin{equation} \label{eq:biconvex}
		\text{minimize} \left\{f(\vect{x},\vect{y}) \mid (\vect{x},\vect{y})\in S \right\}
		\end{equation}
	is defined as a \emph{biconvex problem}, if the feasible set $S$ is biconvex on $X \times Y$ and the objective function $f$ is biconvex on $S$. 		 
	\end{definition}
	
	\begin{theorem}\label{thm:biconvex}
		The optimization problem \eqref{eq:min_mse} is biconvex. 
	\end{theorem}
	\begin{IEEEproof}
		First, we show that $\sum_{b=0}^{B-1}{i_b^2 t_b} \le \mathcal{E}$ is a biconvex set. Note that $i_b^2 t_b$ is a convex function of $i_b$ for every fixed $t_b$. In addition, $i_b^2 t_b$ is a convex function of $t_b$ for every fixed $i_b$, i.e., $\sum_{b=0}^{B-1}{i_b^2 t_b} \le \mathcal{E}$ is a biconvex set. It is clear that $\exp(-2(i_b - 1)t_b)$ is a biconvex function of $i_b$ and $t_b$. Since the positive weight $4^b$ preserves convexity, the objective function is biconvex. In addition, the other constraints are convex. Hence, the problem \eqref{eq:min_mse} is biconvex. 
	\end{IEEEproof}

	Since \eqref{eq:min_mse} is a biconvex problem, we can find suboptimal solutions via \emph{alternate convex search (ACS)}~\cite{Gorski2007biconvex,Wendell1976minimization}. ACS alternatively updates variables by fixing one subset of them and solving the corresponding convex optimization problem. We propose Algorithm~\ref{algo:acs} to optimize the write-current $\vect{i}$ and the write-duration $\vect{t}$. 
	
	\begin{algorithm}
		\caption{ACS algorithm to solve \eqref{eq:min_mse}} \label{algo:acs}
		\begin{algorithmic}[1]
			\State Choose a starting point $\vect{i}^{(0)}$ from the feasible set $S$ and set $k = 0$. 
			\State For a fixed $\vect{i}^{(k)}$, find $\vect{t}^{(k+1)}$ by solving the following convex problem:
			\begin{equation}
			\begin{aligned} \label{eq:min_mse_t}
			& \underset{\vect{t}}{\text{minimize}}
			& & \sum_{b=0}^{B-1}{4^b \exp\left(-2\left(i_b^{(k)} - 1\right)t_b\right)}  \\
			&{\text{subject~to}} & &  \sum_{b=0}^{B-1}{(i_b^{(k)})^2 t_b} \le \mathcal{E} \\
			& & & t_b \ge 0, \quad b=0,\ldots,B-1 \\
			\end{aligned}
			\end{equation}
			\State For a fixed $\vect{t}^{(k+1)}$, find $\vect{i}^{(k+1)}$ by solving the following convex problem. 
			\begin{equation}
			\begin{aligned} \label{eq:min_mse_i}
			& \underset{\vect{i}}{\text{minimize}}
			& & \sum_{b=0}^{B-1}{4^b \exp\left(-2(i_b - 1) t_b^{(k+1)} \right)}  \\
			&{\text{subject~to}} & &  \sum_{b=0}^{B-1}{i_b^2 t_b^{(k+1)}} \le \mathcal{E} \\
			& & & i_b \ge 1 + \epsilon, \quad b=0,\ldots,B-1
			\end{aligned}
			\end{equation}
			\State If the point $(\vect{i}^{(k+1)}, \vect{t}^{(k+1)})$ satisfies a stopping criterion, then stop. Otherwise, set $k:=k+1$ and go back to line 2. 
		\end{algorithmic}
	\end{algorithm}

	\begin{remark}[Starting Point]
		Since biconvex optimization problems may have a large number of local minima~\cite{Gorski2007biconvex}, a starting point $\vect{i}^{(0)}$ can affect the final solution. We empirically find that $\vect{i}^{(0)} = (2, \ldots, 2)$ is a good starting point that minimizes the uniform write-failure probability (see Lemma~\ref{thm:min_wfp}). Numerical evaluation shows that this starting point can achieve comparable results to complicated global nonlinear programming solvers (see Section~\ref{sec:numerical}).  		 
	\end{remark}

	\begin{remark}[Stopping Criterion~\cite{Gorski2007biconvex}]
		There are several ways to define a stopping criterion in Algorithm~\ref{algo:acs}. For example, we can consider the absolute values of the differences between $(\vect{i}^{(k)}, \vect{t}^{(k)})$ and $(\vect{i}^{(k+1)}, \vect{t}^{(k+1)})$ or the difference between $\mathsf{MSE}(\vect{i}^{(k)}, \vect{t}^{(k)})$ and $\mathsf{MSE}(\vect{i}^{(k+1)}, \vect{t}^{(k+1)})$. Alternatively, we can set a maximum number of iterations as a stopping criterion. 
	\end{remark}

	\section{Proposed Iterative Water-filling Algorithm} \label{sec:analysis}

	In this section, we analytically derive the optimal solutions of the individual convex problems \eqref{eq:min_mse_t} and \eqref{eq:min_mse_i} used by the ACS algorithm. Since these problems are convex, we exploit the structure of the problems to derive the optimal solutions analytically using the KKT conditions. By leveraging the derived optimal solutions, we can simplify Algorithm~\ref{algo:acs}.  	
	
	\subsection{Optimal Solutions for Write-Duration}
	
	In this subsection, we present the optimal solutions for the write-duration. First, we consider the optimization problem \eqref{eq:min_mse_t}, which can be interpreted as the classical \emph{water-filling}~\cite{Shannon1949communication}. Afterward, we investigate the optimization problem \eqref{eq:min_mse_t} with the additional latency constraint $\mathsf{L}(\vect{t}) \le \delta$. In this case, the optimization problem can be interpreted as \emph{cave-filling}~\cite{Gao2008optimal,Naidu2016efficient}.
		
	\subsubsection{Water-filling}
	
	The optimal solution of \eqref{eq:min_mse_t} is derived as follows. 	
	\begin{theorem}\label{thm:min_mse_t}
		For a fixed $\vect{i}^{(k)} = (i_0, \ldots, i_{B-1})$, the optimal $\vect{t}^{(k+1)} = \vect{t}^*$ is given by
		\begin{align} \label{eq:min_mse_sol_t}
		t_b^* =   	
		\begin{cases}
		0, & \text{if}\: \log{\nu'} \le \log{\frac{i_b^2}{2 \cdot 4^b \left(i_b - 1\right)}} ; \\
		\frac{\log{\left(\frac{2 \cdot 4^b \left(i_b - 1\right) \nu'}{i_b^2} \right)}}{2 \left(i_b - 1\right)}, & \text{otherwise}.
		\end{cases} \nonumber 
		\end{align}
		Note that $\nu' = \frac{1}{\nu}$ where $\nu$ is a dual variable of the corresponding KKT conditions. 
	\end{theorem}
	\begin{IEEEproof}
		We define the Lagrangian $L_1(\vect{t}, \nu, \boldsymbol{\lambda})$ associated with problem \eqref{eq:min_mse_t} as
		\begin{align}
		L_1(\vect{t}, \nu, \boldsymbol{\lambda}) & = \sum_{b=0}^{B-1}{4^b \exp(-2(i_b - 1)t_b)} \nonumber \\
		&+ \nu \left( \sum_{b=0}^{B-1}{i_b^2 t_b} - \mathcal{E}\right) - \sum_{b=0}^{B-1}{\lambda_b t_b }, 
		\end{align}
		where $\nu$ and $\boldsymbol{\lambda}=(\lambda_0,\ldots,\lambda_{B-1})$ are the dual variables. 
	\end{IEEEproof}

\begin{figure}[t]
		\centering
		\vspace{-3mm}
		\includegraphics[width=0.45\textwidth]{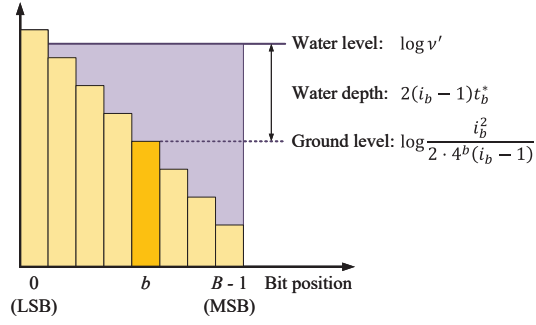}
		\vspace{-4mm}
		\caption{Water-filling interpretation of Theorem~\ref{thm:min_mse_t}.}
		\label{fig:wf}
	\end{figure}

	The optimal solution of Theorem \ref{thm:min_mse_t} can be interpreted as classical \emph{water-filling} as shown in Fig.~\ref{fig:wf}. Each bit position can be regarded as an individual channel among $B$ parallel channels. The ground levels depend on the importance of the bit positions, i.e., the more significant bit position corresponds to the lower ground level. Hence, larger durations are assigned to more significant bit positions. For a bit position $b$ such that $\log{\nu'} > \log{\frac{i_b^2}{2 \cdot 4^b \left(i_b - 1\right)}}$, $t_b^*$ satisfies the following condition (see Appendix~\ref{pf:min_mse_t}): 	
	\begin{equation} \label{eq:wf}
		\log{\nu'} = \log{\frac{i_b^2}{2 \cdot 4^b (i_b - 1)}} + 2(i_b - 1) t_b^*
	\end{equation}
	where $\log{\nu'}$, $\log{\frac{i_b^2}{2 \cdot 4^b (i_b - 1)}}$, and $2(i_b - 1) t_b^*$ represent the water level, the ground level, and the water depth, respectively. 
	
	The water level $\log{\nu'}$ depends on the energy budget $\mathcal{E}$. For a bit position whose ground level is higher than the water level (as in the LSB of Fig.~\ref{fig:wf}), the optimal duration is zero, which means that this bit position would not be written.
	
	\subsubsection{Cave-filling}
	
	If we impose an additional constraint on latency, i.e., $\mathsf{L}(\vect{t}) \le \delta$, then the optimal solution of \eqref{eq:min_mse_t} is derived as follows. 

	\begin{theorem}\label{thm:min_mse_t_cf}
		For a fixed $\vect{i}^{(k)} = (i_0, \ldots, i_{B-1})$, the optimal $\vect{t}^{(k+1)} = \vect{t}^*$ is given by
		\begin{align}
			&t_b^* =   		\nonumber \\
			&\begin{cases}
				0, & \text{if}\: \log{\nu'} \le \log{\frac{i_b^2}{2 \cdot 4^b \left(i_b - 1\right)}} ; \\
				\delta, & \text{if}\: \log{\nu'} \ge \log{\frac{i_b^2}{2 \cdot 4^b \left(i_b - 1\right)}} + 2(i_b - 1)\delta; \\
				\frac{\log{\left(\frac{2 \cdot 4^b \left(i_b - 1\right) \nu'}{i_b^2} \right)}}{2 \left(i_b - 1\right)}, & \text{otherwise}.
			\end{cases} \nonumber 
		\end{align}
	\end{theorem}
	\begin{IEEEproof}
		We can merge $t_b \ge 0$ and $\mathsf{L}(\vect{t}) \le \delta$ into $0 \le t_b \le \delta$ for $b \in [0,B-1]$ where $[0,B-1]$ denotes $\{0, \ldots, B-1\}$. Then, the corresponding Lagrangian is given by
		\begin{align}
			L_2(\vect{t}, \nu, \boldsymbol{\lambda}) & = \sum_{b=0}^{B-1}{4^b \exp(-2(i_b - 1)t_b)} \nonumber \\
			&+ \nu \left( \sum_{b=0}^{B-1}{i_b^2 t_b} - \mathcal{E}\right) - \sum_{b=0}^{B-1}{\lambda_b t_b } \nonumber \\
			&+ \sum_{b=0}^{B-1}{\eta_b (t_b - \delta)},	
		\end{align}
		where $\nu$, $\boldsymbol{\lambda}=(\lambda_0,\ldots,\lambda_{B-1})$, and $\boldsymbol{\eta}=(\eta_0,\ldots,\eta_{B-1})$ are the dual variables. The details of the proof are given in Appendix~\ref{pf:min_mse_t_cf}. 
	\end{IEEEproof}

	\begin{figure}[t]
	\centering
	\vspace{-3mm}
	\includegraphics[width=0.45\textwidth]{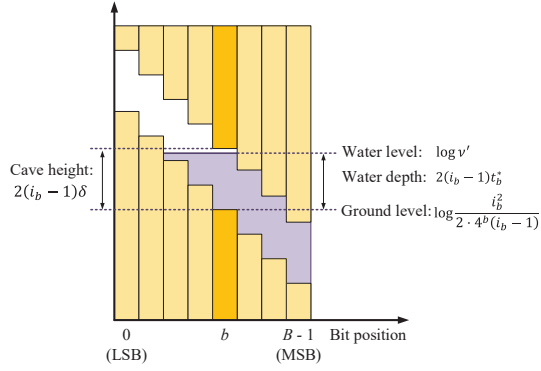}
	\vspace{-4mm}
	\caption{Cave-filling interpretation of Theorem~\ref{thm:min_mse_t_cf}.}
	\label{fig:cf}
	\end{figure}	

	The optimal solution of Theorem~\ref{thm:min_mse_t_cf} can be interpreted as \emph{cave-filling} as shown in Fig.~\ref{fig:cf}. Because of the latency constraint, the water depth cannot be greater than $2(i_b - 1)\delta$, which corresponds to the cave height. We note that cave-filling interpretation is similar to \emph{water-filling problem under peak power constraints} in wireless communications~\cite{Gao2008optimal,Naidu2016efficient}.

	\subsection{Optimal Solutions for Write-Current}
	
	\begin{theorem}\label{thm:min_mse_i}
		For a fixed $\vect{t}^{(k+1)} = \vect{t}$, the optimal $\vect{i}^{(k+1)} = \vect{i}^*$ of \eqref{eq:min_mse_i} is given by
		\begin{equation}\label{eq:min_mse_sol_i}
		i_b^* =
		\begin{cases}
		1 + \epsilon, & \text{if}\: \mu \ge \frac{4^b}{1 + \epsilon} e^{-2t_b \epsilon}  ; \\
		\frac{1}{2t_b}W\left(\frac{2\cdot 4^b t_b e^{2t_b}}{\mu}\right), & \text{otherwise}
		\end{cases}
		\end{equation}
		where $\mu$ is a dual variable. Also, $W(\cdot)$ denotes the \emph{Lambert W function} (i.e., the inverse function of $f(x) = xe^x$)~\cite{Corless1996lambert}. 
	\end{theorem}
	\begin{IEEEproof}
			We define the Lagrangian $L_3(\vect{i}, \mu, \boldsymbol{\lambda})$ associated with problem \eqref{eq:min_mse_i} as
		\begin{align}
		L_3(\vect{i}, \mu, \boldsymbol{\lambda}) & = \sum_{b=0}^{B-1}{4^b e^{-2(i_b - 1)t_b}} + \mu \left( \sum_{b=0}^{B-1}{i_b^2 t_b} - \mathcal{E}\right) \nonumber \\
		& - \sum_{b=0}^{B-1}{\lambda_b \left\{i_b - (1 + \epsilon)\right\} } 	
		\end{align}
		where $\mu$ and $\boldsymbol{\lambda} = (\lambda_0,\ldots,\lambda_{B-1})$ are the dual variables. The details of the proof are given in Appendix~\ref{pf:min_mse_i}.		
	\end{IEEEproof}
	Since $W(x)$ is an increasing function for $x \ge 0$, the larger currents are assigned to more significant bit positions.

	\subsection{Proposed Iterative Water-filling Algorithm}
	
	By leveraging the derived optimal solutions, we propose an iterative water-filling algorithm. The proposed algorithm (Algorithm~\ref{algo:acs_simple}) can find identical solutions to the original ACS (Algorithm~\ref{algo:acs}), but with much less computational effort. 
	
	We note that the dual variables $\nu'$ (in Theorem~\ref{thm:min_mse_t} and Theorem~\ref{thm:min_mse_t_cf}) and $\mu$ (in Theorem~\ref{thm:min_mse_i}) depend on the write energy budget $\mathcal{E}$. They can be efficiently obtained by the bisection method as in~\cite{Palomar2005practical}.  
	
	\begin{algorithm}
		\caption{Proposed iterative water-filling algorithm to solve \eqref{eq:min_mse}} \label{algo:acs_simple}
		\begin{algorithmic}[1]
			\State Choose a starting point $\vect{i}^{(0)}$ from the feasible set $S$ and set $k = 0$. 
			\State For a fixed $\vect{i}^{(k)}$, set $\vect{t}^{(k+1)}$ by Theorem~\ref{thm:min_mse_t} (or Theorem~\ref{thm:min_mse_t_cf} with a latency constraint). 
			\State For a fixed $\vect{t}^{(k+1)}$, set $\vect{i}^{(k+1)}$ by Theorem~\ref{thm:min_mse_i}. 
			\State If the point $(\vect{i}^{(k+1)}, \vect{t}^{(k+1)})$ satisfies a stopping criterion, then stop. Otherwise, set $k:=k+1$ and go back to line 2. 
		\end{algorithmic}
	\end{algorithm}

	Originally, the authors of~\cite{Yu2004iterative} proposed an iterative water-filling-type algorithm to maximize the sum capacity for a Gaussian vector multiple-access channel. It decomposes the multiuser problem into a sequence of identical single-user problems. On the other hand, our iterative water-filling algorithm decomposes the write optimization problem into a sequence of write-current optimization problem and write-duration optimization problem. Furthermore, our optimization problem is non-convex whereas maximizing the sum capacity of Gaussian vector multiple-access channel is convex.

	
	\section{Convergence and MSE Reduction} \label{sec:convergence}
	
	In this section, we prove \emph{convergence} of the proposed iterative water/cave-filling algorithm. Furthermore, we show that the propose algorithm can reduce the MSE \emph{exponentially} with $B$. 

	\subsection{Convergence of Proposed Algorithm}

	We show that Algorithm~\ref{algo:acs_simple} guarantees convergence to a locally optimal MSE. The converged MSE depends on a starting point. 

	\begin{lemma}\label{thm:monotone}The sequence $\left\{\mathsf{MSE}(\vect{i}^{(k)}, \vect{t}^{(k)})\right\}_{k \in \mathbb{N}}$ obtained by Algorithm~\ref{algo:acs_simple} is monotonically decreasing, i.e., $\mathsf{MSE}(\vect{i}^{(k+1)}, \vect{t}^{(k+1)}) \le \mathsf{MSE}(\vect{i}^{(k)}, \vect{t}^{(k)})$ for all $k \in \mathbb{N}$. 
	\end{lemma}
	\begin{IEEEproof} Note that $\mathsf{MSE}(\vect{i}^{(k)}, \vect{t}^{(k+1)}) \le \mathsf{MSE}(\vect{i}^{(k)}, \vect{t}^{(k)})$ and $\mathsf{MSE}(\vect{i}^{(k+1)}, \vect{t}^{(k+1)}) \le \mathsf{MSE}(\vect{i}^{(k)}, \vect{t}^{(k+1)})$ because of \eqref{eq:min_mse_t} and \eqref{eq:min_mse_i}, respectively. Hence, $\mathsf{MSE}(\vect{i}^{(k+1)}, \vect{t}^{(k+1)}) \le \mathsf{MSE}(\vect{i}^{(k)}, \vect{t}^{(k)})$. 
	\end{IEEEproof}
	
	\begin{theorem}\label{thm:convergence}The sequence $\left\{\mathsf{MSE}(\vect{i}^{(k)}, \vect{t}^{(k)})\right\}_{k \in \mathbb{N}}$ obtained by Algorithm~\ref{algo:acs_simple} converges monotonically. 
	\end{theorem}
	\begin{IEEEproof} It is clear that $\mathsf{MSE}(\vect{i}^{(k)}, \vect{t}^{(k)}) \ge 0$ for all $k \in \mathbb{N}$ by \eqref{eq:proxy} and \eqref{eq:mse}. Then, $\left\{\mathsf{MSE}(\vect{i}^{(k)}, \vect{t}^{(k)})\right\}_{k \in \mathbb{N}}$ is monotonically decreasing and bounded from below, and thus $\left\{\mathsf{MSE}(\vect{i}^{(k)}, \vect{t}^{(k)})\right\}_{k \in \mathbb{N}}$ converges because of monotone convergence theorem. 
	\end{IEEEproof}

%

	\subsection{MSE Reduction of Proposed Algorithm}
	
	In this subsection, we show that the proposed Algorithm~\ref{algo:acs_simple} can reduce the MSE exponentially with $B$ if the latency constraint is not imposed. 
	
	Suppose that the starting point is $\vect{i}^{(0)} = (2, \ldots, 2)$. By Theorem~\ref{thm:min_mse_t}, Algorithm~\ref{algo:acs_simple} provides the following optimized write-durations $\widetilde{\vect{t}} = (\widetilde{t}_0^*, \ldots, \widetilde{t}_{B-1}^*)$ where 
	\begin{equation}\label{eq:min_mse_sol_t_i0}
	\widetilde{t}_b^* =
	\begin{cases}
	0, & \text{if}\: \nu \ge \frac{4^b}{2} ; \\
	\frac{1}{2}\log{\left(\frac{1}{\nu}\cdot \frac{4^b }{2} \right)}, & \text{otherwise}.
	\end{cases}
	\end{equation}
	
	\begin{lemma}\label{thm:positive_t}
		If $\mathcal{E} > 2B(B-1)\log{2}$, then $\widetilde{t}_b > 0$ for all $b \in [0, B-1]$ and 
		\begin{equation} \label{eq:positive_t}
		\widetilde{t}_b = \frac{\mathcal{E}}{4B} + \left(b - \frac{B-1}{2}\right) \cdot \log{2}. 
		\end{equation}  
	\end{lemma}
	\begin{IEEEproof}
	The proof is given in Appendix~\ref{pf:mse_improvement}. 
	\end{IEEEproof}

	\begin{theorem}\label{thm:mse_improvement}
		If $\mathcal{E} > 2B(B-1)\log{2}$, then Algorithm~\ref{algo:acs_simple} can achieve the following MSE reduction ratio $\gamma$:
		\begin{align} 
		\gamma &= \frac{ \mathsf{MSE}\left(\vect{i}^*, \vect{t}^* \right)} {\mathsf{MSE} \left(\vect{i}^{(0)}, \vect{t}^{(0)} \right) } \le \frac{ \mathsf{MSE}\left(\vect{i}^{(0)}, \widetilde{\vect{t}} \right)} {\mathsf{MSE} \left(\vect{i}^{(0)}, \vect{t}^{(0)} \right) } \nonumber \\
		& = \frac{3B}{2} \cdot \frac{ 2^B}{4^{B}- 1} \approx \frac{3B}{2} \cdot 2^{-B} \label{eq:mse_improvement}
		\end{align}
		where $\mathsf{MSE}(\vect{i}^{(0)}, \widetilde{\vect{t}} )$  is given by
		\begin{equation}
		\mathsf{MSE}\left(\vect{i}^*, \vect{t}^* \right) = c' \cdot \frac{B}{2} \cdot 2^{B} \exp\left(- \frac{\mathcal{E}}{2B}\right).
		\end{equation}		
		In addition, $\mathsf{MSE} (\vect{i}^{(0)}, \vect{t}^{(0)} )$ (i.e., the MSE by uniform energy allocation) is given by
		\begin{equation}
		\mathsf{MSE} \left(\vect{i}^{(0)}, \vect{t}^{(0)} \right) = c' \cdot \frac{4^B - 1}{3} \exp\left(-\frac{\mathcal{E}}{2B}\right)	\end{equation}		
		where $\vect{i}^{(0)}$ and $\vect{t}^{(0)}$ are the uniform values to satisfy the energy constraint. 
	\end{theorem}
	\begin{IEEEproof}
		The proof is given in Appendix~\ref{pf:mse_improvement}. 
	\end{IEEEproof}
	Note that $\mathsf{MSE} \left(\vect{i}^{(0)}, \vect{t}^{(0)} \right)$ is the MSE corresponding to the parameters minimizing the write-failure probability for the uniform-significance case (see Lemma~\ref{thm:min_wfp}).
	
	We showed that Algorithm~\ref{algo:acs_simple} can reduce the MSE exponentially with $B$, compared to the parameters optimized for uniform write-failure probability. Although we cannot guarantee global optimality of the proposed algorithm, in the following section, we empirically show that the proposed algorithm achieves comparable MSE to complicated nonlinear programming solvers.

	\section{Numerical Results}\label{sec:numerical}
	
	We evaluate the optimized solutions for the write-failure probability for single-bits as well as the MSE for $B$-bit words. The critical current $I_c$ and the characteristic relaxation time $T_c$ do not affect the numerical results because the normalized values $i = \tfrac{I}{I_c}$ and $t =\frac{T}{T_c}$ are considered. As in \cite{Khvalkovskiy2013basic}, we set $\Delta = 60$ for the thermal stability factor.  
	
\begin{figure}[t]
	\centering
	\includegraphics[width=0.45\textwidth]{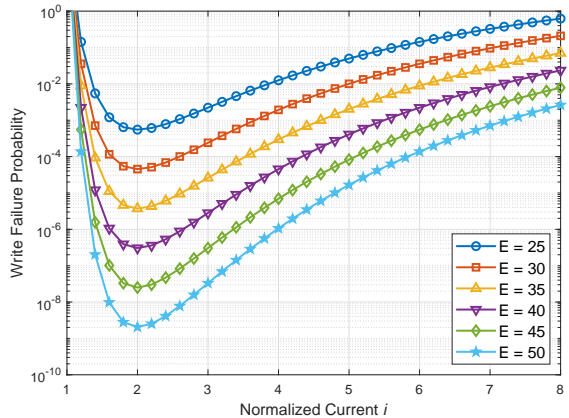}
	\caption{Normalized write-current to minimize the write-failure probability (see Lemma~\ref{thm:min_wfp}) for several values of the energy constraint.}
	\label{fig:p_opt}
	\end{figure}

\begin{figure}[t!]
	\centering
	\subfloat[]{\includegraphics[width=0.45\textwidth]{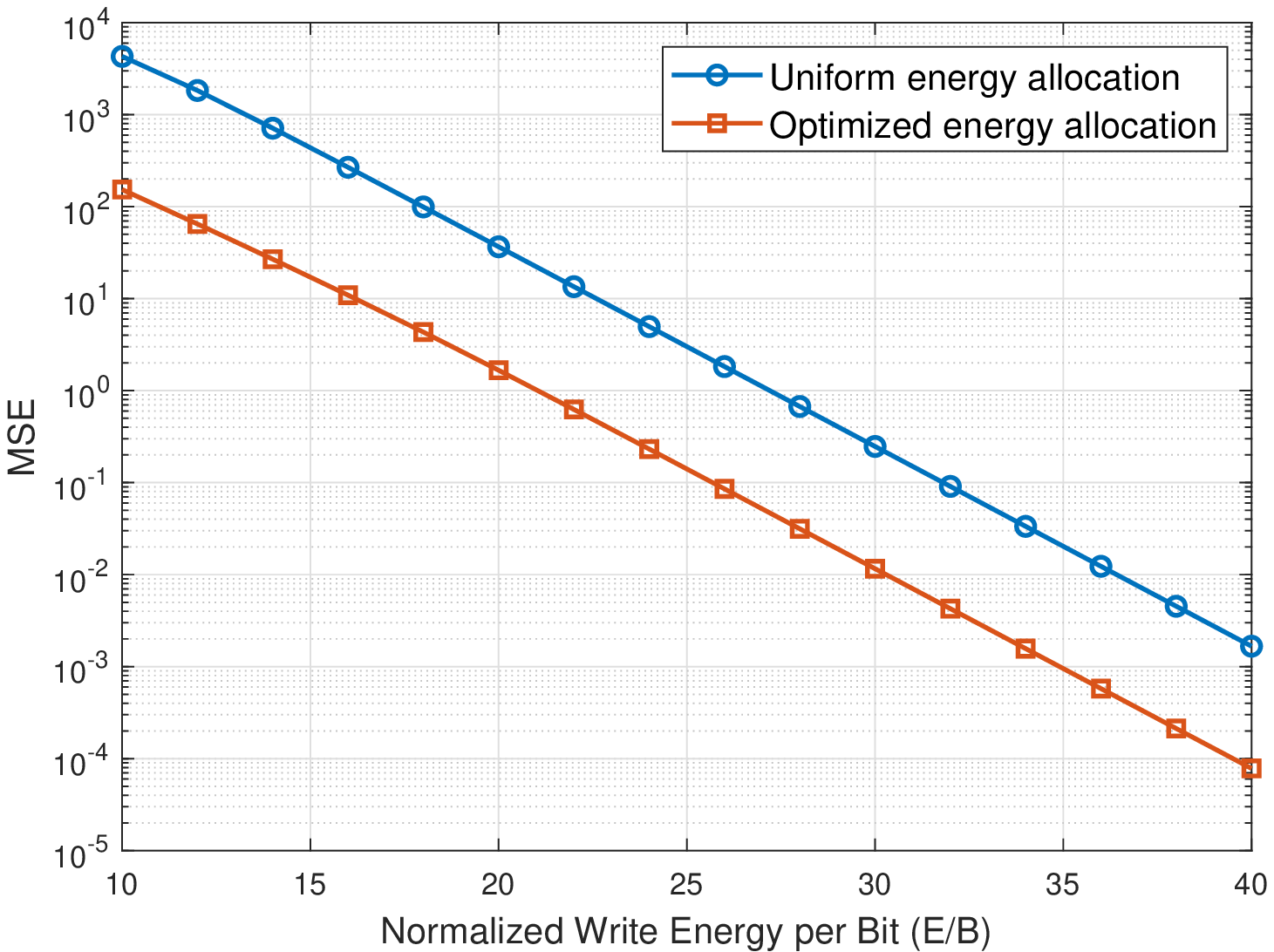}
		\label{fig:optimal_mse}}
	\hfil
	\subfloat[]{\includegraphics[width=0.45\textwidth]{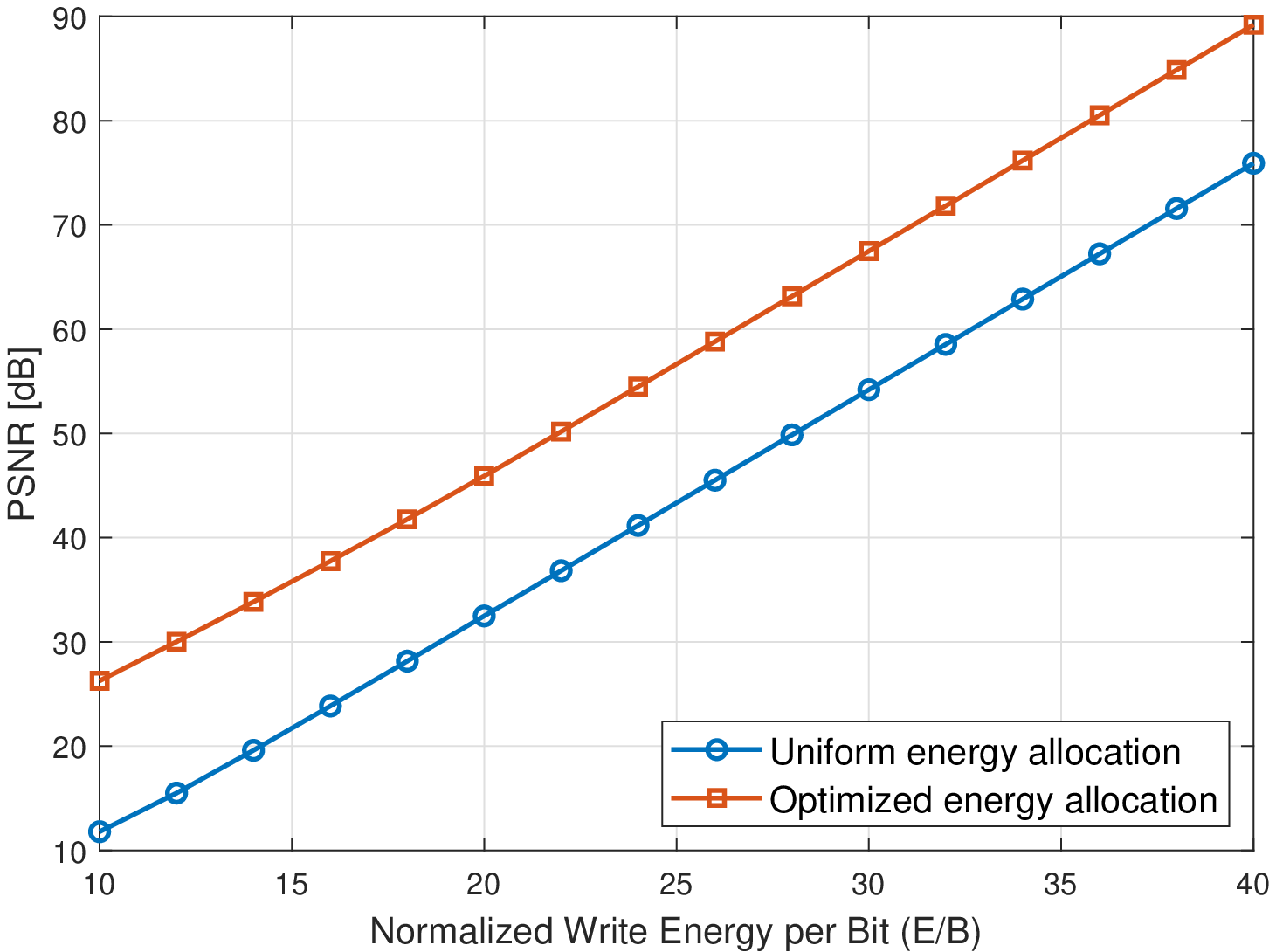}
		\vspace{-4mm}
		\label{fig:optimal_psnr}}
	\caption{Comparison of the conventional uniform energy allocation and the optimized energy allocation by Algorithm~\ref{algo:acs} ($B = 8$): (a) MSE and (b) PNSR.}
	\label{fig:optimal}
	\end{figure}

\begin{figure}[t!]
	\centering
	\includegraphics[width=0.45\textwidth]{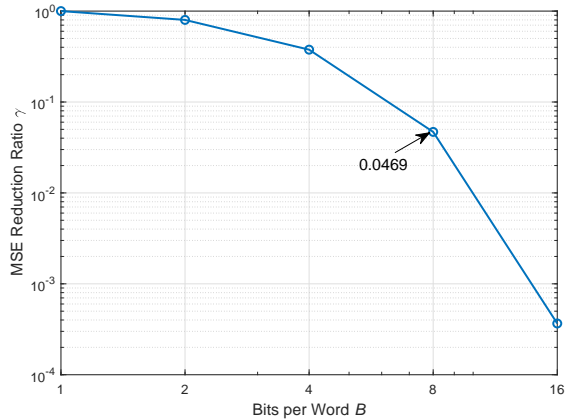}
	\caption{The MSE reduction ratio $\gamma$ by Theorem~\ref{thm:mse_improvement}.}
	\label{fig:mse_improvement}
	\end{figure}

\begin{figure}[t!]
	\centering
	\includegraphics[width=0.45\textwidth]{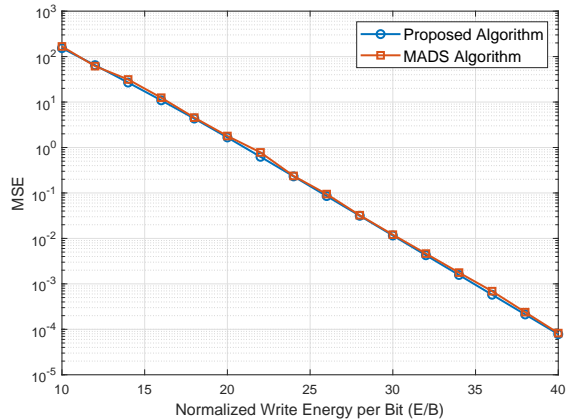}
	\caption{MSE comparison of the proposed algorithm and MADS algorithm.}
	\label{fig:mse_global}
	\end{figure}

\begin{figure}[t!]
		\centering
		\includegraphics[width=0.45\textwidth]{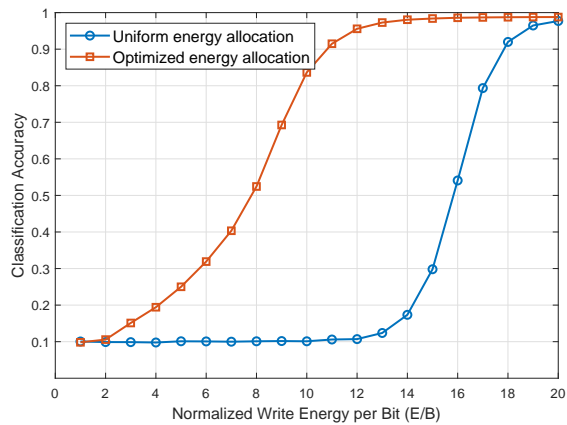}
		\caption{MNIST classification accuracy.}
		\label{fig:mnist}
	\end{figure}

	We provide the following numerical results. 
	\begin{itemize}
		\item Minimizing the write-failure probability (Fig.~\ref{fig:p_opt}); 
		\item Minimizing the MSE and peak signal-to-noise ratio (PSNR) (Figs.~\ref{fig:optimal} and~\ref{fig:mse_improvement}); 
		\item MSE comparison of the proposed algorithm and complicated nonlinear solvers (Fig.~\ref{fig:mse_global});
		\item MNIST classification results by deep neural networks (Fig.~\ref{fig:mnist}). 
	\end{itemize}

	First, Fig.~\ref{fig:p_opt} shows that $i^* = 2$ and $t^* = \frac{\mathcal{E}}{4}$ minimize the write-failure probability for single-bit write operation (as proved in Lemma~\ref{thm:min_wfp}). The corresponding minimal write-failure probability decreases exponentially with the write energy $\mathcal{E}$ as shown in \eqref{eq:p_opt}. 
	
	Fig.~\ref{fig:optimal} compares the uniform energy allocation (optimized for single-bit write operation) and the optimized write energy allocation (optimized for multi-bit write operation). Fig.~\ref{fig:optimal}\subref{fig:optimal_mse} compares the MSEs of the uniform energy allocation and the optimized energy allocation by Algorithm~\ref{algo:acs_simple}. As predicted by Theorem~\ref{thm:mse_improvement}, the MSE reduction ratio is $\gamma \approx \frac{3B}{2}\cdot2^{-B} = 0.0469$ for $B = 8$, which corresponds to the improvement by a factor of 21. 
	
	Fig.~\ref{fig:optimal}\subref{fig:optimal_psnr} compares the PSNRs, which is a widely used fidelity metric for image and video quality. The PSNR depends on the MSE as $\mathsf{PSNR} = 10 \log_{10}{\frac{(2^B-1)^2}{\mathsf{MSE}}}$. At $\mathsf{PSNR} = \SI{40}{dB}$, the optimized write energy allocation can reduce the write energy by \SI{24}{\%}. 

	Fig.~\ref{fig:mse_improvement} shows that the MSE reduction ratio improves exponentially with $B$ (as derived in Theorem~\ref{thm:mse_improvement}). Although we cannot guarantee the optimality, the proposed Algorithm~\ref{algo:acs} is very effective in reducing the MSE. Note that $\gamma \approx  0.0469$ for $B=8$ and $\gamma \approx 3.66 \times 10^{-4}$ for $B = 16$. 
	
	We compare the MSE results of Algorithm~\ref{algo:acs_simple} to the mesh adaptive direct search (MADS) algorithm~\cite{Audet2006mesh}. The numerical results by MADS algorithm are obtained by NOMAD, which is a widely used software package for global nonlinear programming~\cite{LeDigabel2011algorithm}. Fig.~\ref{fig:mse_global} shows that the MSEs obtained by the proposed algorithm and NOMAD are almost identical. Note that the proposed algorithm is much more computationally efficient.

	We conduct simulations on the MNIST dataset for handwritten character recognition. We used the 8-bit quantized pretrained neural network of~\cite{Sakr2017analytical}. The neural network architecture is given by 784--512--512--512--10, i.e., 3 hidden layers (a detailed description can be found in~\cite{Sakr2017analytical}). We assume that the quantized model parameters are stored in MRAM cells and these models suffer from bit-flipping errors due to MRAM's write failures. 
	
	Fig.~\ref{fig:mnist} shows that the optimized energy allocation by the proposed algorithm can improve the classification accuracy significantly. Equivalently, the optimized energy allocation can achieve the same classification accuracy with much less energy. To achieve \SI{90}{\%} of classification accuracy, the optimized write energy allocation requires 11 normalized energy units per bit whereas the uniform write energy allocation requires 18 normalized energy units per bit, i.e., the proposed algorithm achieves \SI{40}{\%} energy reduction.

	\section{Conclusion}\label{sec:conclusion}
	
	We addressed MRAM's write energy optimization problem, which is a major challenge of MRAM technology. After formulating the biconvex optimization problem, we proposed the iterative water-filling algorithm to minimize the MSE under a write energy budget. Also, we proved that the proposed algorithm converges and the optimized solutions can reduce the MSE exponentially. Numerical evaluation supports that the proposed algorithm effectively improves PSNR (for image processing applications) and classification accuracy (for machine learning applications). 
	

	\appendices
	
	\section{Proof of Proposition~\ref{lem:proxy}}\label{pf:proxy}
	From~\eqref{eq:wfp}, 
	\begin{align}
		p_{\text{WF}}(i,t) &= 1 - \exp\left(-\frac{\Delta \pi^2 (i-1)}{4\left\{i \exp(2(i-1)t) - 1 \right\}}\right) \nonumber \\
		&\simeq \frac{\Delta \pi^2 (i-1)}{4\left\{i \exp(2(i-1)t) - 1 \right\}} \label{eq:pf_proxy_s1} \\
		&\simeq \frac{\Delta \pi^2}{4}\cdot\frac{i-1}{i}\cdot \exp\left(-2(i-1)t \right) \label{eq:pf_proxy_s2} \\		
		&\simeq \frac{\Delta \pi^2}{4} \exp\left(-2(i-1)t \right) \label{eq:pf_proxy_s3}
	\end{align}
	where \eqref{eq:pf_proxy_s1} follows from Taylor approximation. Also, \eqref{eq:pf_proxy_s2} and \eqref{eq:pf_proxy_s3} hold for sufficiently large $i$ and $t$. 
	
	\section{Proof of Lemma~\ref{thm:min_wfp}}\label{pf:min_wfp}
	
	It is clear that $i^*$ and $t^*$ satisfy $i^2 t = \mathcal{E}$ to maximize $(i-1)t$. Then, we can set $t = \frac{\mathcal{E}}{i^2}$ and the corresponding objective function is given by
	\begin{equation}
	g(i) = (i-1)t = \mathcal{E} \cdot \frac{i-1}{i^2}.  
	\end{equation} 
	Since $g'(i) = \mathcal{E} \cdot \frac{2-i}{i^3}$, $g'(2) = 0$ and $g'(i) < 0$ for $i > 2$. Hence, $g(i)$ is maximized when $i^* = 2$ and $t^* = \frac{\mathcal{E}}{4}$. 
		
	\section{Proof of Theorem~\ref{thm:min_mse_t}}\label{pf:min_mse_t}
	
	The corresponding KKT conditions are as follows:
	\begin{align}
	\sum_{b=0}^{B-1}{i_b^2 t_b} &\le \mathcal{E}, \quad \nu \ge 0, \quad
	\nu \cdot \left(\sum_{b=0}^{B-1}{
		i_b^2 t_b} - \mathcal{E}\right) = 0, \label{eq:cr1_KKT_t_1} \\
	t_b &\ge 0, \quad \lambda_b \ge 0, \quad \lambda_b t_b = 0 \label{eq:cr1_KKT_t_2} \\
	\frac{\partial L_1}{\partial t_b} &= -2 \cdot 4^b(i_b-1) e^{-2(i_b - 1)t_b} + \nu i_b^2 - \lambda_b = 0 \label{eq:cr1_KKT_t_3}
	\end{align}
	for $b \in [0, B-1]$. From \eqref{eq:cr1_KKT_t_3}, $\lambda_b$ is given by
	\begin{equation} \label{eq:cr1_KKT_t_lambda}
	\lambda_b = i_b^2 \left(\nu  - \frac{2 \cdot 4^b(i_b-1)}{i_b^2} \cdot e^{-2(i_b - 1)t_b} \right). 
	\end{equation}
	Suppose that $\nu = 0$. Then $\lambda_b < 0$ because of $i_b \ge 1 + \epsilon$, which violates the condition of $\lambda \ge 0$. Hence, $\nu \ne 0$ and 
	\begin{equation} \label{eq:cr1_KKT_energy}
	\sum_{b=0}^{B-1}{i_b^2 t_b} = \mathcal{E}.
	\end{equation}   
		
	From \eqref{eq:cr1_KKT_t_2} and \eqref{eq:cr1_KKT_t_lambda}, 
	\begin{align}
	\lambda_b t_b  = i_b^2 \cdot t_b \left\{\nu  - \frac{2 \cdot 4^b(i_b-1)}{i_b^2} \cdot e^{-2(i_b - 1)t_b} \right\} = 0. \label{eq:cr1_KKT_t_slack_1}
	\end{align}
	Because of $\lambda_b \ge 0$ and $i_b \ge 1 + \epsilon$, we obtain 
	\begin{equation} \label{eq:cr1_KKT_t_nu}
	\nu \ge \frac{2 \cdot 4^b(i_b-1)}{i_b^2} \cdot e^{-2(i_b - 1)t_b}. 
	\end{equation}	
		
	If $\nu \ge \frac{2 \cdot 4^b(i_b-1)}{i_b^2}$, then $t_b = 0$. Otherwise (i.e., $t_b > 0$ and $\lambda_b = 0$), then $\nu = \frac{2 \cdot 4^b(i_b-1)}{i_b^2} \cdot e^{-2(i_b - 1)t_b}$. Because of $e^{-2(i_b - 1)t_b} < 1$ for $t_b > 0$, which contradicts to $\nu \ge \frac{2 \cdot 4^b(i_b-1)}{i_b^2}$. 
	
	If $\nu < \frac{2 \cdot 4^b(i_b-1)}{i_b^2}$, then $t_b = 0$ is not allowed because of \eqref{eq:cr1_KKT_t_nu}. Hence, $t_b > 0$ and $\lambda_b = 0$. By \eqref{eq:cr1_KKT_t_lambda} and $i_b \ge 1+\epsilon$,
	\begin{equation} \label{eq:cr1_KKT_t_nu_eq}	
	\nu = \frac{2 \cdot 4^b(i_b-1)}{i_b^2} \cdot e^{-2(i_b - 1)t_b},  
	\end{equation}
	which results in
	\begin{equation} \label{eq:pf_wf_opt}
	t_b^* = \frac{1}{2(i_b - 1)} \log\left(\frac{1}{\nu} \cdot \frac{2\cdot4^b (i_b - 1)}{i_b^2}\right). 
	\end{equation}	
	We note that \eqref{eq:pf_wf_opt} satisfies the following condition: 
	\begin{equation} \label{eq:pf_wf}
	\log{\frac{1}{\nu}} = \log{\frac{i_b^2}{2 \cdot 4^b (i_b - 1)}} + 2(i_b - 1) t_b. 
	\end{equation}

	\section{Proof of Theorem~\ref{thm:min_mse_t_cf}}\label{pf:min_mse_t_cf}

	The corresponding KKT conditions are as follows:
	\begin{align}
		\sum_{b=0}^{B-1}{i_b^2 t_b} &\le \mathcal{E}, \quad \nu \ge 0, \quad
		\nu \cdot \left(\sum_{b=0}^{B-1}{
			i_b^2 t_b} - \mathcal{E}\right) = 0, \label{eq:cr1_KKT_t_cf_1} \\
		t_b &\ge 0, \quad \lambda_b \ge 0, \quad \lambda_b t_b = 0 \label{eq:cr1_KKT_t_cf_2} \\
		t_b &\le \delta, \quad \eta_b \ge 0, \quad \eta_b (t_b - \delta) = 0 \label{eq:cr1_KKT_t_cf_3} \\
		\frac{\partial L_2}{\partial t_b} = -2 \cdot & 4^b (i_b-1) e^{-2(i_b - 1)t_b} + \nu i_b^2 - \lambda_b + \eta_b = 0 \label{eq:cr1_KKT_t_cf_4}
	\end{align}
	for $b \in [0, B-1]$. From \eqref{eq:cr1_KKT_t_cf_4}, $\eta_b$ is given by
	\begin{equation} \label{eq:cr1_KKT_t_cv_lambda}
		\eta_b = 2 \cdot 4^b (i_b-1) e^{-2(i_b - 1)t_b} - \nu i_b^2 + \lambda_b. 
	\end{equation}

	Suppose that $\nu = 0$. Then, $\eta_b = 2 \cdot 4^b (i_b-1) e^{-2(i_b - 1)t_b} + \lambda_b > 0 $ because of $i_b \ge 1 + \epsilon$ and $\lambda_b \ge 0$. Since $\eta_b (t_b - \delta) = 0$, $\nu = 0$ leads to $t_b = \delta$ for $b \in [0, B-1]$. Since $\vect{t} = (\delta, \ldots, \delta)$ is a trivial solution, we claim that $\nu \ne 0$ and $\sum_{b=0}^{B-1}{
		i_b^2 t_b} = \mathcal{E}$.
	
	From $\lambda_b \ge 0$ and $\eta_b \ge 0$, we obtain
	\begin{equation} \label{eq:cr1_KKT_t_cf_5}
		\nu i_b^2 - \lambda_b \le 2 \cdot 4^b (i_b - 1) e^{-2(i_b - 1)t_b} \le \nu i_b^2 + \eta_b. 
	\end{equation}
	
	By \eqref{eq:cr1_KKT_t_cf_2} and \eqref{eq:cr1_KKT_t_cf_3}, the optimal solution satisfies one of the following conditions: 1) $t_b = 0$, $\lambda_b > 0$, and $\eta_b = 0$; 2) $t_b = \delta$, $\lambda_b = 0$, and $\eta_b > 0$; and 3) $0 \le t_b \le \delta$ and $\lambda_b = \eta_b = 0$. 
	
	
	1) $t_b = 0$, $\lambda_b > 0$, and $\eta_b = 0$
	
	By $\eta_b = 0$ and \eqref{eq:cr1_KKT_t_cf_5}, we obtain the condition $\log{\nu'} \le \log{\frac{i_b^2}{2 \cdot 4^b \left(i_b - 1\right)}}$ where $\nu' = \frac{1}{\nu}$. 
	
	2) $t_b = \delta$, $\lambda_b = 0$, and $\eta_b > 0$
	
	By $\lambda_b = 0$, $t_b = \delta$, and \eqref{eq:cr1_KKT_t_cf_5}, we obtain the condition $\log{\nu'} \ge \log{\frac{i_b^2}{2 \cdot 4^b \left(i_b - 1\right)}} + 2(i_b - 1) \delta$. 
	
	3) $0 \le t_b \le \delta$ and $\lambda_b = \eta_b = 0$
	
	By setting $\lambda_b = \eta_b = 0$ in \eqref{eq:cr1_KKT_t_cf_5}, we obtain
	\begin{equation}
		\nu i_b^2 = 2 \cdot 4^b (i_b - 1) e^{-2(i_b - 1)t_b}, 
	\end{equation}
	which is the optimal $t_b^* = \frac{\log{\left(\frac{2 \cdot 4^b \left(i_b - 1\right) \nu'}{i_b^2} \right)}}{2 \left(i_b - 1\right)}$.

	\section{Proof of Theorem~\ref{thm:min_mse_i}}\label{pf:min_mse_i}

	The corresponding KKT conditions are as follows:
	\begin{align}
	\sum_{b=0}^{B-1}{i_b^2 t_b} &\le \mathcal{E}, \quad \mu \ge 0, \quad
	\mu \cdot \left(\sum_{b=0}^{B-1}{
		i_b^2 t_b} - \mathcal{E}\right) = 0, \label{eq:cr1_KKT_i_1} \\
	i_b &\ge 1 + \epsilon, \quad \lambda_b \ge 0, \quad \lambda_b \{i_b - (1 + \epsilon)\}  = 0 \label{eq:cr1_KKT_i_2} \\
	\frac{\partial L_3}{\partial i_b} &= -2 \cdot 4^b t_b e^{-2(i_b - 1)t_b} + 2 \mu t_b i_b  - \lambda_b = 0 \label{eq:cr1_KKT_i_3}
	\end{align}
	for $b \in [0, B-1]$. From \eqref{eq:cr1_KKT_i_3}, $\lambda_b$ is given by
	\begin{equation} \label{eq:cr1_KKT_i_lambda}
	\lambda_b = 2 t_b i_b \left(\mu - \frac{4^b e^{- 2t_b (i_b - 1)}}{i_b} \right).
	\end{equation}
	Suppose that $\mu = 0$. Then, $\lambda_b = -2t_b \cdot 4^b e^{-2t_b(i_b -1 )} \le 0$, which is true only if $t_b = 0$ for all $b \in [0, B-1]$. Since this is a trivial case, we focus on $\mu \ne 0$ and $\sum_{b=0}^{B-1}{i_b^2 t_b} = \mathcal{E}$.
	
	If $t_b = 0$, then the corresponding $i_b$ affects neither the MSE nor the energy. Hence, we suppose that $t_b \ne 0$. If $\lambda_b = 0$, then 
	\begin{equation}\label{eq:cr1_KKT_i_nu}
		\mu = \frac{4^b e^{-2t_b (i_b - 1)}}{i_b},
	\end{equation} 
	which is equivalent to
	\begin{equation}\label{eq:cr1_KKT_i_lambert}
		\frac{2\cdot 4^b t_b e^{2t_b}}{\mu} = 2t_b i_b e^{2t_b i_b} = z e^z
	\end{equation}
	where $z = 2t_b i_b$. Then, $z = 2t_b i_b = W\left(\frac{2\cdot 4^b t_b e^{2t_b}}{\mu}\right)$ where $W(\cdot)$ denotes the Lambert W function~\cite{Corless1996lambert}.  
	Hence, 
	\begin{equation} \label{eq:cr1_KKT_i_opt1}
	i_b^* = \frac{1}{2t_b} W\left(\frac{2\cdot 4^b t_b e^{2t_b}}{\mu}\right). 
	\end{equation}
	Note that $i_b^* = 1+ \epsilon$ for $\mu = \frac{4^b e^{-2t_b \epsilon}}{1 + \epsilon}$ because of $W(ze^z) = z$.   
		
	Suppose that $g(i_b) = \frac{4^b e^{-2t_b (i_b - 1)}}{i_b}$ in \eqref{eq:cr1_KKT_i_lambda}. Because of $\frac{\text{d}g(i_b)}{\text{d}i_b} < 0$, $g(i_b)$ is a monotonically decreasing function. If $\mu > \frac{4^b e^{-2t_b \epsilon}}{1 + \epsilon} = g(1 + \epsilon)$, then $\lambda_b \ne 0$ and $i_b^* = 1 + \epsilon$ by \eqref{eq:cr1_KKT_i_2}.

%

	\section{Proof of Lemma~\ref{thm:positive_t} and Theorem~\ref{thm:mse_improvement}}\label{pf:mse_improvement}
	
	From \eqref{eq:min_mse_sol_t_i0}, we observe that $\widetilde{t}_b > 0$ for all $b \in [0, B-1]$ if $\nu < \frac{1}{2}$. By \eqref{eq:cr1_KKT_energy}, $(\vect{i}^{(0)}, \widetilde{\vect{t}} )$ satisfies
	\begin{equation}
	\sum_{b=0}^{B-1}{4 t_b} = \sum_{b=0}^{B-1}{2 \log{\left(\frac{1}{\nu}\cdot \frac{4^b}{2}\right)}} = \mathcal{E}, 
	\end{equation}
	which results in 
	\begin{equation}\label{eq:pf_nu}
	\nu = 2^{B-2} \cdot \exp\left( -\frac{\mathcal{E}}{2B} \right). 
	\end{equation}
	Then, the condition $\nu < \frac{1}{2}$ is equivalent to
	\begin{equation} \label{eq:pf_energy_condition}
	\mathcal{E} > 2B(B-1) \log{2}. 
	\end{equation}
	Hence, $\widetilde{t}_b > 0$ for all $b \in [0, B-1]$ 
	if \eqref{eq:pf_energy_condition} holds. By \eqref{eq:min_mse_sol_t_i0} and \eqref{eq:pf_nu}, we obtain \eqref{eq:positive_t}. 
	
	By \eqref{eq:mse} and \eqref{eq:positive_t}, 
	\begin{align}
	\mathsf{MSE}\left(\vect{i}^{(0)}, \widetilde{\vect{t}}\right) & = c  \cdot \sum_{b=0}^{B-1}{4^b \exp\left(-2 \widetilde{t}_b\right)} \nonumber \\
	& = c \cdot B \cdot 2^{B-1} \exp\left(- \frac{\mathcal{E}}{2B}\right). \label{eq:pf_mse_nonuniform}
	\end{align}	
	The uniform energy allocation of $(\vect{i}^{(0)}, \vect{t}^{(0)})$ results in 
	\begin{equation}
	\mathsf{MSE}\left(\vect{i}^{(0)}, \vect{t}^{(0)}\right) = c \cdot \frac{4^B - 1}{3} \exp\left(-\frac{\mathcal{E}}{2B}\right). \label{eq:pf_mse_uniform}
	\end{equation}
	From \eqref{eq:pf_mse_nonuniform} and \eqref{eq:pf_mse_uniform}, we obtain \eqref{eq:mse_improvement}.

	
	\section*{Acknowledgment}
	
	This work was partly supported by the DGIST Start-up Fund Program of the Ministry of Science and ICT 2021010014. The work of Yuval Cassuto was partly supported by the US-Israel Binational Science Foundation, and by the Israel Science Foundation.
	
	
	
	\bibliographystyle{IEEEtran}
	\bibliography{abrv,mybib}

\end{document}